\documentclass[prl,preprint,amsmath,amssymb,showpacs]{revtex4}
\usepackage[english]{babel}
\usepackage[applemac]{inputenc}
\usepackage{latexsym}

\newcommand{\sig}{$\sigma_1(\omega)$}
\newcommand{\ea}{\emph{et al.~}}

\usepackage{hyperref}
\usepackage{graphicx}
\usepackage{epstopdf}

\newif\ifpdf
\ifx\pdfoutput\undefined
\pdffalse 
\else
\pdfoutput=1 
\pdftrue
\fi

\usepackage{graphicx}

\usepackage{xspace}
\usepackage{subfigure}

\begin{document}


\title {Optical evidence for a magnetically driven structural transition in the spin web $Cu_3TeO_6$}

\ifpdf
\DeclareGraphicsExtensions{.jpg,.pdf}
\else
\DeclareGraphicsExtensions{.eps,.jpg}
\fi

\author {G. Caimi}
\author {L. Degiorgi}
\affiliation{Laboratorium f\"ur Festk\"orperphysik, ETH
Z\"urich,
CH-8093 Z\"urich, Switzerland}\
\author {H. Berger}
\author{L. Forr\'o}
\affiliation{Institut de physique de la mati\`ere complexe (IPMC), EPF Lausanne, CH-1015 Lausanne, Switzerland}\

\date{\today}

\begin{abstract}
$Cu_3TeO_6$ is a modest frustrated $S=1/2$ spin system, which undergoes an anti-ferromagnetic transition at $T_N\sim61$ $K$. The anti-ferromagnetic spin alignment in $Cu_3TeO_6$ below $T_N$ is supposed to induce a magneto-elastic strain of the lattice. The complete absorption spectrum of $Cu_3TeO_6$ is obtained through Kramers-Kronig transformation of the optical reflectivity, measured from the far-infrared up to the ultraviolet spectral range as a function of temperature ($T$). Below $T^*\sim 50$ $K$, we find a new  mode at 208 $cm^{-1}$. The spectral weight associated to this additional mode increases as  $\propto (T^*-T)^{1/2}$ with decreasing $T$ below $T^*$. The implication of the optical findings will be discussed in relation to the magnetic phase transition at $T_N$.
\end{abstract}


\maketitle
Following the investigation of the high temperature superconducting cuprates and the search for related transition-metal oxides, a fascinating field of copper oxide compounds, vanadates, manganates and nickelates opened up. These compounds show effects of strong electronic correlations and magnetism in low dimensions, in particular the interplay between spins, charge and orbital degrees of freedom \cite{lemmens}. Of paramount interest is the study of magnetism relying on $3d^9$ $Cu^{2+}$ ions, which reveals a manifold of magnetic structures and variegated physical properties. Such a diversity in the physics of $3d$ $Cu^{2+}$ systems principally originates from their effective magnetic dimensionality. Furthermore, the level of frustration and the important quantum fluctuations in these compounds may lead to either singlet non-magnetic, disordered spin liquid or magnetically long-range ordered ground states at low temperatures.

In that context, $Cu_3TeO_6$ is of relevance, since a novel type of magnetic lattice, referred to as a three dimensional spin web, has been identified \cite{herak}. In this lattice, almost planar neighboring $Cu^{2+}$ hexagons share one common corner and, by buckling and folding in space, form a complex three dimensional network. The spin susceptibility $\chi_{spin}$ follows the Curie-Weiss law \cite{herak} between 170 and 330 $K$ with Curie-Weiss temperature $\Theta_{CW}\sim -145$ $K$. Large and negative $\Theta_{CW}$ suggests that $Cu$-spins are strongly antiferromagnetically coupled. Below 170 $K$, $\chi_{spin}$ increases less rapidly than the Curie-Weiss law and then decreases below $T_N=61$ $K$. Magnetic peaks appear in the neutron powder diffraction experiment below $T_N$. Their integrated intensity varies as the square of the $S=1/2$ Brillouin function with decreasing temperature \cite{herak}. The results of neutron diffraction studies are indeed consistent with a collinear antiferromagnetic (AF) alignment of the spins within the hexagons. The competition between the local anisotropy of the hexagons and AF nearest neighbors interaction leads to a modest frustration ($f=2.4$), which is resolved below $T_N$ by the formation of the AF state \cite{herak}. In passing, we note that the magnetic properties of $Cu_3TeO_6$ share some common features with those of the 3D pyrochlore lattice \cite{lee,ramirez}. In these latter systems, however, the more pronounced frustration relies on the geometrically frustrated spin tetrahedra building blocks.

The spins in the collinear AF ordered phase are aligned along the $\lbrack111\rbrack$ direction \cite{herak}. Such an alignment of spins is supposed to induce a magneto-elastic strain, which could lead to a structural phase transition. Neutron diffraction powder experiments are not conclusive on this issue. Low temperature X-ray diffraction and new neutron scattering experiments on a single domain crystal would be of extreme interest in order to clarify the consequence of the magnetic state on the structural properties. Here, we offer another approach in order to further shed light on the intrinsic physical properties of $Cu_3TeO_6$. We perform optical reflectivity experiments and extract the complete absorption spectrum. Besides the excitations related to the electronic band structure, the absorption spectrum in the infrared spectral range is of particular relevance, since it reveals information on the lattice dynamics (phonon modes). We find a peculiar excitation at about 208 $cm^{-1}$, which appears below 50 $K$ and whose intensity grows in a mean-field like fashion for decreasing temperature. This behaviour could suggest the development of a novel mode, possibly induced by a magneto-striction effect.

Single crystals of $Cu_3TeO_6$ were grown using $CuTeO_3-TeO_2$ flux \cite{herak}. Polycrystalline $CuTeO_3$ was mixed with an excess of $TeO_2$ and heated in a covered aluminia crucible up to 800 $^0C$ for about 24 hours. After slowly cooling to room temperature platelet-like single crystals of $Cu_3TeO_6$ could be mechanically separated from the flux. Our $Cu_3TeO_6$ specimens have a size of $2.5\times 4.3 $ mm. The great quality of our sample is further confirmed by the measurement of the magnetic susceptibility, displaying the antiferromagnetic transition below 63 $K$ \cite{bill}. We measure the optical reflectivity $R(\omega)$ from the far-infrared (5 $meV$) up to the ultraviolet (12 $eV$), as a function of temperature ($T$) and magnetic field. We did not find any magnetic field dependence in our spectra. The absorption spectrum represented by the real part $\sigma_1(\omega)$ of the optical conductivity is obtained through Karmers-Kronig (KK) transformation of $R(\omega)$. Since $Cu_3TeO_6$ is an insulator at all temperatures, $R(\omega)$ was extrapolated with a constant value for $\omega\to 0$. Above the highest measurable frequency, $R(\omega)$ was extended into the electronic continuum with the standard extrapolations $R(\omega)\approx \omega^{-s}$ ($2<s<4$). Details pertaining to the experiment can be found in Refs. \onlinecite{wooten} and \onlinecite{dressel}.

\begin{figure} [!h]
\begin{center}
\resizebox{9.0 cm}{!}{\includegraphics{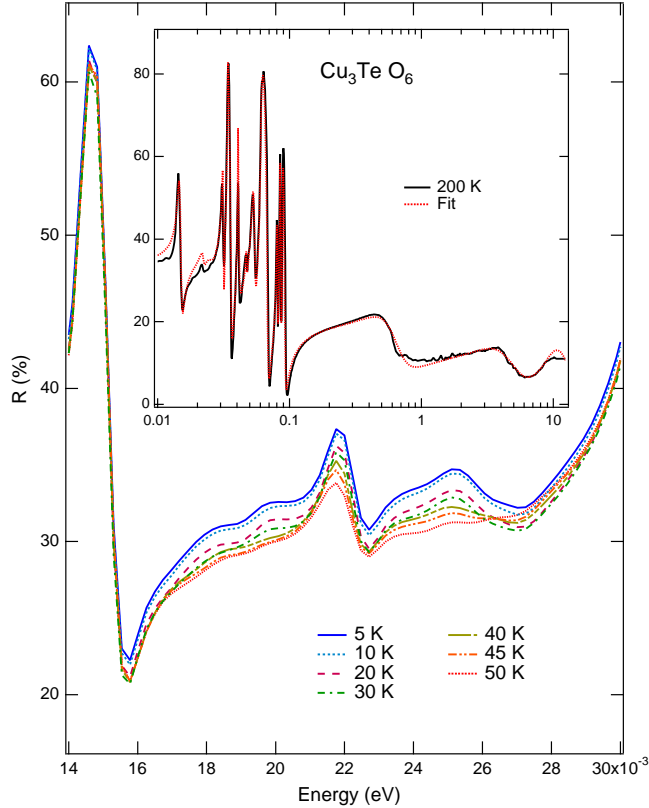}}
\caption{(Color online) $R(\omega)$ spectra of  $Cu_3TeO_6$~in   the spectral range from 14 to 30 $meV$. We just illustrate the curves between 5 and 50 $K$, where the temperature dependence is observed. The inset shows $R(\omega)$ at 200 $K$ and its phenomenological Lorentz fit (see text).}
   \label{rs}  
\end{center}
\end{figure}

\begin{figure} [!h]
\begin{center}
    \resizebox{9.0 cm}{!}{\includegraphics{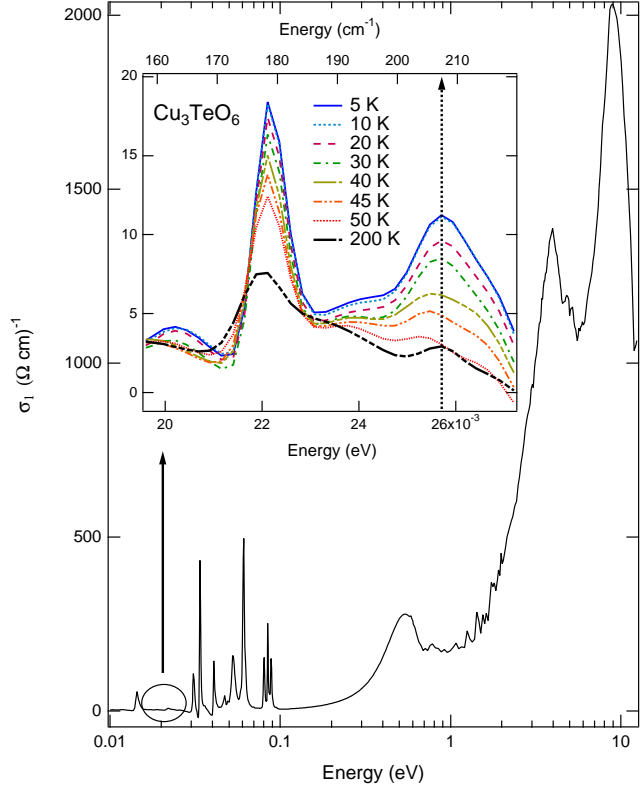}}
 \caption{(Color online) \sig~of $Cu_3TeO_6$~at 200 $K$. The inset is a blow-up of the region where the new mode at 208 $cm^{-1}$ (arrow) develops below about 50 $K$.  }
\label{sig}
\end{center}
\end{figure}

Figure \ref{rs} shows the temperature dependence of $R(\omega)$ in the energy interval between 14 and 30 $meV$ and for temperatures below 50 $K$. At high temperatures, all spectra look pretty much similar, without relevant differences within the experimental error.  As an example for the high-$T$ reflectivity, we plot $R(\omega)$ measured at 200 $K$ in the inset of Fig. \ref{rs}. The real part $\sigma_1(\omega)$ of the optical conductivity is illustrated in Fig. \ref{sig}. We first remark the overall insulating nature of the $Cu_3TeO_6$~system, with \sig~tending to zero in the $dc$ limit, at all temperatures.  Furthermore, the absorption spectra below 800 $cm^{-1}$ ($\sim0.1$ $eV$) are characterized by an array of sharp absorptions, detected  precisely at 119, 179, 208 (only below 50 $K$),  251, 274, 327, 379, 427, 491, 648, 681 and 713 $cm^{-1}$.  

Group analysis considerations allow us to predict the number of phonon modes expected in $Cu_3TeO_6$. The $Cu_3TeO_6$~crystal has a cubic structure and belongs to the space group $Ia3$, which is isomorphic to the point group $T_h$ (Ref. \onlinecite{falk}). The Bravais  space cell is composed by four formula units with three inequivalent oxygen positions. The site symmetry for the three inequivalent $O$ sites, which accommodate eight atoms each,  is $C_3$. The four $Te$ atoms have  $C_{3i}$ symmetry, while the twelve $Cu$ atoms have a $C_2$ symmetry \cite{falk}.  Therefore, using the correlation methods \cite{cm}, we estimate  the complete phonon spectrum as:
\begin{eqnarray}
\Gamma^{cryst}&=&3\Gamma^{C_3}+\Gamma^{C_2}+\Gamma^{C_{3i}}\nonumber\\
&=&4A_g+4E_{g} +14F_{g}+5A_u+5E_{u}+17F_{u}.\nonumber
\end{eqnarray}
After subtracting the acoustic   mode, namely $\Gamma^{acustic}= F_{u}$,  one obtains for the  Raman and  infrared (IR) active modes:
\begin{eqnarray}
\Gamma^{Raman}&=&4A_g+4E_{g} +14F_{g} \nonumber \\
\Gamma^{IR}&=&16 F_{u}. \nonumber
 \end{eqnarray}
 In our \sig~spectra, we detect  only eleven modes at high temperatures,   less than the  group theory predictions. This is of no surprise since some phonon modes might have an equivalent energy and therefore are not  detectable separately. Alternatively, their intensity might be below the detection threshold for weak dipole displacements.  
 
The  \sig~spectrum at high frequencies is characterized by three main absorptions at 0.55, 3.8 and 9.2 $eV$, which are ascribed to electronic interband transitions. We assume that the $3d$ $Cu$ bands of the $Cu_3TeO_6$ spin system are located around the Fermi energy. Since the $Cu$ atoms are surrounded by a distorted octahedral arrangement \cite{falk}, one expects a split of the $3d$ bands at least in a $t_{2g}$- and a $e_g$-complex. A further split of these two sub-bands is also possible because of the lowering of symmetry, of electron-electron interaction or of Jahn-Teller effect. Therefore, we propose to attribute the feature at 0.55 $eV$  to a transition within the different $3d$ bands. The well developed absorptions at 3.8 $eV$ and at 9.2 $eV$ are assigned to interband transitions from the $O$ $2p$ and the $Te$  $4p$ states into the $Cu$ $3d$ bands. Of course, band structure calculations would be of great help, in order to unambiguously identify these transitions.  

\begin{figure} [!h]
\begin{center}
    \resizebox{7.0 cm}{!}{\includegraphics{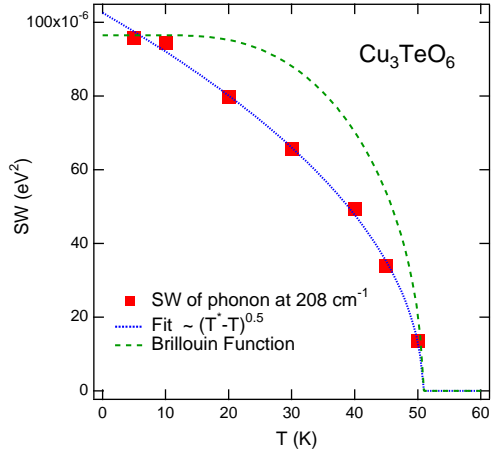}}
    \caption{(Color online) Temperature dependence of the spectral weight ($SW$) of the additional mode which develops at 208 $cm^{-1}$ below $T^*=50$ $K$. The dashed line refers to a calculation with a $S=1/2$ Brillouin function, while the dotted line is the $(T^*-T)^{0.5}$ power law. }
     \label{ad-ph}
  \end{center}
\end{figure}

We have also applied the phenomenological Lorentz fit to $R(\omega)$ and $\sigma_1(\omega)$ \cite{wooten,dressel}, in order to catch the main features of the absorption spectrum and its temperature dependence. A Lorentz harmonic oscillator has been ascribed to each absorption features (phonon modes and electronic interband transitions). The same set of fit components well reproduces both $R(\omega)$ and $\sigma_1(\omega)$ spectra at each temperature. The great fit quality is exemplified by the fit of $R(\omega)$ at 200 $K$ in the inset of Fig. \ref{rs}. The temperature dependence of the \sig~spectra is moderate, except for the spectral range between 14 and 30 $meV$, which displays the appearance of the new mode at 208 $cm^{-1}$ below $T^*\sim 50$ $K$. This is undoubtedly demonstrated by the inset of Fig. \ref{sig}. The intensity of this additional mode increases continuously for $T<T^*$ and  saturates approximately at 10 $K$. In order to appreciate the peculiar temperature dependence of the mode at 208 $cm^{-1}$ below $T^*$, it is worth noting that the adjacent mode at 179 $cm^{-1}$ progressively narrows with decreasing temperature already below 200 $K$. The phenomenological fit of the mode at 208 $cm^{-1}$ with a Lorentz harmonic oscillator allows us to extract the mode strength as well as the mode damping. The damping is found to be constant at each temperature while the  oscillator strength increases by lowering $T$. Of particular relevance for the present discussion is the spectral weight ($SW$) encountered in the mode at 208 $cm^{-1}$, which is obtained either by squaring the strength of the Lorentz harmonic oscillator or by integrating \sig~in the energy interval between 195 and 220 $cm^{-1}$ (inset Fig. 2). Both procedures are totally equivalent. The relative change of $SW$ is plotted in Fig. \ref{ad-ph}, which was obtained by first subtracting a background, represented by the average amount of $SW$ in the same energy interval and at temperatures above 50 $K$. The resulting temperature dependence of $SW$ follows quite closely the square root behaviour $(T^*-T)^{0.5}$, as demonstrated in Fig. \ref{ad-ph}. It is quite astonishing that the square root behaviour is found to be valid down to very low temperatures. This temperature dependence also differs from the behaviour seen in the integrated intensity of the magnetic peaks appearing below $T_N$ in the neutron diffraction experiment \cite{herak}. Indeed, it is evident from Fig. 3 that the $S=1/2$ Brillouin-function does not  account for the temperature dependence of $SW$ observed in our spectra around 208 $cm^{-1}$, unless very close to $T^*$.

The new mode, developing at 208 $cm^{-1}$ below $T^*$, might originate from a reduction of the crystal symmetry, which induces the activation of a new infrared phonon mode. While a local distortion can not be excluded a priori, the development of the mode at 208 $cm^{-1}$ most probably implies the occurrence of a structural transition at $T<T^*$, associated to the AF ordering below $T_N$ \cite{herak}. Indeed, Herak \ea sustain that a structural transition should be the direct consequence of the magneto-elastic strain, associated to the collinear AF alignment of the spins along the [111] direction \cite{herak}. 
However, it is puzzling, that the temperature $T^*$ for the onset of the mode at 208 $cm^{-1}$ is smaller than $T_N$ \cite{nota}. A possible way for solving this issue might reside in the fact that the intensity of the peaks in the neutron diffraction experiments reaches approximately 2/3 of their saturation level at about 50 $K$. Therefore, we might speculate that only below $T^*$ the necessary magnetization level has been reached in order to drive the structural transition. 

It is worth to point out that a similar feature has been found at slightly higher energy (at about 400 $cm^{-1}$) with decreasing temperature in several multilayered cuprates \cite{basovrmp}, like the underdoped $YBa_2Cu_3O_{7-x}$ (Ref. \onlinecite{homes}) as well as in the ladder $Sr_2Ca_{12}Cu_{24}O_{41}$ (Ref. \onlinecite{ruzicka}) system, along the out-of-plane direction. Its macroscopic origin is still debated and ranges from transverse plasma resonance to Josephson currents \cite{basovrmp}. Nevertheless, this feature, which incidentally appears at an energy scale comparable to the spin gap \cite{dagotto} of the ladders, was also associated to a phonon anomaly due to a change in the lattice dynamics (e.g., activation of silent or Raman active phonons), as consequence of the symmetry breaking.

Magnetic excitations are in principle not infrared active but can under favourable conditions carry a finite electric dipole moment, as it has been demonstrated for the $S=1/2$ Heisenberg spin chain $\alpha'-NaV_2O_5$ \cite{damascelli}. Moreover, Lorenzana and Sawatzky proposed a model based on a phonon assisted activation of the electric dipole moment of a two-magnon excitation \cite{lorenzana}, in order to account for the multimagnon sidebands discovered in undoped copper oxides by infrared transmission spectroscopy \cite{perkins}. Nevertheless, the low energy scale associated with the mode at 208 $cm^{-1}$ as well as the absence of any magnetic field dependence in our experiment make these latter possibilities quite unlikely.

In conclusion, a new mode at 208 $cm^{-1}$ develops in the absorption spectrum of $Cu_3TeO_6$ at $T<50$ $K$. Our optical data strongly support the presence of a structural transition driven by the AF ordering of the spins below $T_N$. This emphasizes the important role played by the spin-phonon coupling. An overall mean-field like behaviour governs the temperature dependence of the strength of this new mode. Nevertheless, it remains to be seen how one can reconcile the different temperature dependence of the spectral weight encountered in the mode at 208 $cm^{-1}$ and of the integrated intensity of the magnetic peaks in the neutron diffraction experiment. Our results are also challenging from the theoretical point of view; there is no obvious mechanism of the spin-Peierls type which, given the proposed spin structure \cite{herak}, would lead to a gain in magnetic energy. This is very much at variance with the typical scenario in spin 1/2 chains or in some frustrated magnets \cite{lemmens}.

\begin{acknowledgments}
The authors wish to thank B. Pedrini for the sample characterization, J. M\"uller for technical help, and  O. Zaharko, P. Lemmens, F. Mila and A.
Sacchetti for fruitful discussions. This work
has been supported by the Swiss National Foundation for the
Scientific Research within the NCCR MaNEP pool.
\end{acknowledgments}


\begin{thebibliography}{10}
\bibitem{nota}
The fact, that $T_N$ in our sample is of about 63 $K$ \cite{bill}, excludes the possibility to eventually ascribe $T^*$ with an effective Neel temperature, reduced because of $Cu$ vacancies.
\end{thebibliography}

\end{document}